\documentclass[
 amsmath,amssymb,
preprint,%
nofootinbib
]{revtex4}

\usepackage{bm}
\usepackage{makecell}
\usepackage{multirow}
\usepackage{graphicx}
\usepackage{epstopdf}
\usepackage{tablefootnote}
\usepackage{subfigure}
\usepackage{amsmath}
\usepackage{color}
\usepackage{lscape}

\usepackage{amssymb}
\usepackage{amsfonts}
\usepackage{mathrsfs}
\usepackage{graphicx,subfigure,booktabs}
\usepackage{verbatim}
\usepackage{ulem}
\usepackage{setspace}
\usepackage{url}
\usepackage[utf8]{inputenc}
\usepackage{float}

\usepackage[colorlinks,
linkcolor=blue,
filecolor=blue,
anchorcolor=blue,
urlcolor=blue,
citecolor=blue,
bookmarks=false,
]{hyperref}

\usepackage{fancyhdr}

\begin{document}


\title{Correlating $B\to K^{(\ast)} \nu\bar{\nu}$ and flavor 
anomalies in SMEFT}

\author{Feng-Zhi Chen}
\author{Qiaoyi Wen}
\author{Fanrong Xu}
\email{fanrongxu@jnu.edu.cn}
\affiliation{Department of Physics, College of Physics $\&$ Optoelectronic Engineering, Jinan University, Guangzhou 510632, P.R. China}

\date{\today}
\small

\begin{abstract}

The recent measurement of $\mathcal{B}(B^+\to K^+\nu\bar{\nu})$ by Belle-II reveals a $2.8~\sigma$ deviation from the Standard Model (SM) prediction. Combining this with a prior Belle measurement of $\mathcal{B}(B^{0}\to K^{\ast0}\nu\bar{\nu})$, the upper bound of the ratio $\mathcal{B}(B^{0}\to K^{\ast0}\nu\bar{\nu})/\mathcal{B}(B^+\to K^+\nu\bar{\nu})$ is notably smaller than the SM prediction. In this work, tensions are solved within the framework of Standard Model Effective Field Theory (SMEFT). The flavor-changing-neutral-current (FCNC) and charged-current observables of either down-type ($b\to s\nu\bar{\nu}$, $b\to s\ell^+\ell^-$, and $b\to u_i\ell\bar{\nu}$) or up-type ($u_j\to u_i\nu\bar{\nu}$, $u_j\to u_i\ell^+\ell^-$, and $u_j\to s\ell\bar{\nu}$) processes, described by Low-Energy Effective Field Theory (LEFT) operators, are interconnected by a minimal set of four SMEFT operators at the electroweak scale. Subsequently, we obtain the latest ranges of Wilson coefficients for these four operators through a global fit that accommodates flavor anomalies such as $R_{K^{(\ast)}}$, $R_{D^{(\ast)}}$, and $\mathcal{B}(B\to K^{(\ast)}\nu\bar{\nu})$. Our findings reveal that predictions for $\mathcal{B}(B^+\to \tau^+\nu_\tau)$ and $\mathcal{B}(D_s^+\to \tau^+\nu_\tau)$ align well with measured values from Belle and BESIII, based on the fitted coefficients. The predicted branching fraction for $B^0\to K^{\ast0}\nu\bar{\nu}$ is $(1.42\pm 0.74)\times 10^{-5}$, closely approaching the current experimental upper limit. Anticipation surrounds the rare decay $B_s\to \tau^+ \tau^-$, expected in the near future with a branching fraction on the order of $10^{-4}$.

\end{abstract}

\keywords{Suggested keywords}
\maketitle

\section{Introduction}
\label{sec:intro}
The rare $B$ decays, including neutral-current and charged-current processes, can not only provide us the very clean laboratory to study properties of Quantum Chromodynamics (QCD), but can also be served as the ideal places to search for new physics (NP) beyond the SM. Recently, Belle II collaboration reported a first measurement on the branching ratio of $B^+\to K^+\nu\bar{\nu}$~\cite{Belle-II:2023esi},\footnote{Note that $\mathcal{B}(B^+\to K^+\nu\bar{\nu})$ have also been priorly measured by Belle~\cite{Belle:2013tnz,Belle:2017oht} and BaBar~\cite{BaBar:2010oqg,BaBar:2013npw}, since no statistically significant signals are observed and only upper limits on $\mathcal{B}(B^+\to K^+\nu\bar{\nu})$ at $90\%$ confidence level (CL) are reported in these experiments, we do not combine the Belle II result~\cite{Belle-II:2023esi} with them in our numerical analysis.}
\begin{align}
\mathcal{B}(B^+\to K^+\nu\bar{\nu})_\mathrm{exp}=(2.3\pm0.7)\times 10^{-5}\,,
\end{align}
which has a significance of $3.5\sigma$ with respect to the background-only hypothesis, and is in tension with the SM prediction~\cite{Becirevic:2023aov}
\begin{align}
\mathcal{B}(B^+\to K^+\nu\bar{\nu})_\mathrm{SM}=(4.43\pm0.31)\times 10^{-6}
\end{align}
at $2.8\sigma$.\footnote{The long-distance contribution from $B^+\to \tau^+(\to K^+\bar{\nu}_\tau)\nu_\tau$ has been excluded.} At quark level, $B^+\to K^+\nu\bar{\nu}$ decay shares the same transition $b\to s\nu\bar{\nu}$ with $B^0\to K^{\ast0}\nu\bar{\nu}$ decay, a prior Belle measurement on the latter gives $\mathcal{B}(B^0\to K^{\ast0}\nu\bar{\nu})_\mathrm{exp}<1.8\times10^{-5}$~\cite{Belle:2017oht},\footnote{In the numerical analysis we assume that the experimental input $\mathcal{B}(B^0\to K^{\ast0}\nu\bar{\nu})<1.8\times10^{-5}$ obeys the half-normal distribution, so it can be translated into the formalism with a center value and a symmetric error: $\mathcal{B}(B^0\to K^{\ast0}\nu\bar{\nu})=(0.87\pm0.66)\times10^{-5}$.}
 about twice the SM prediction $\mathcal{B}(B^0\to K^{\ast0}\nu\bar{\nu})_\mathrm{SM}=(9.47\pm1.40)\times10^{-6}$~\cite{Becirevic:2023aov}.
By defining the following ratio:
\begin{align}\label{eq:definition}
R\equiv\frac{\mathcal{B}(B^0\to K^{\ast0}\nu\bar{\nu})}{\mathcal{B}(B^+\to K^+\nu\bar{\nu})}\,,
\end{align}
one easily deduces that the current experiment bound on such a ratio is $R_\mathrm{exp}=0.38\pm0.31$, which is notably smaller than that of the SM prediction $R_\mathrm{SM}=2.14\pm0.35$. This, together with the discrepancy in $\mathcal{B}(B^+\to K^+\nu\bar{\nu})$, has revived a flurry of theoretical explanations by using either heavy or light NP models.  As for the former, the missing energies in the experiment are considered to be taken away only by the SM neutrinos, so the NP, such as scalar leptoquarks~\cite{Becirevic:2018afm}, can contribute to $b\to s\nu\bar{\nu}$ via the effective low-energy operators after the heavy degrees of freedom are integrated out. While for the latter, the missing energies are assumed to be carried by other light NP particles, e.g., the sterile neutrino~\cite{Felkl:2023ayn}, axion~\cite{MartinCamalich:2020dfe}, dark matter~\cite{Filimonova:2019tuy,Datta:2023iln}, etc., so the construction of Lagrangian will depend on the category of light NP particles that involved. Other relevant studies on this aspect see also Refs.~\cite{Athron:2023hmz,Bause:2023mfe,Allwicher:2023xba,He:2023bnk,Chen:2023wpb,Altmannshofer:2023hkn,McKeen:2023uzo,Ho:2024cwk,Chen:2024cll,Bolton:2024egx,Marzocca:2024hua,Hou:2024vyw}. Specifically, we would like to highlight Ref.~\cite{Hou:2024vyw} as one of the most intriguing studies, where the Belle-II excess is interpreted within two new physics scenarios. This study considers both heavy new mediators and light final states, such as dark matter or axion-like particles, within the frameworks of SMEFT and DSMEFT (SMEFT extended by dark matter), respectively.

In this work, we will assume that the NP originate from above the electroweak (EW) scale and, there are no extra new light particles present below the EW scale, so only the SM left-handed neutrinos are involved in the Lagrangian. Since there is no direct evidence indicating the presence of NP particles, a model independent analysis by employing the effective field theory will be a reasonable method. In this way, the $b\to s\nu\bar{\nu}$ transition below the EW scale can be described by the well-known LEFT~\cite{Jenkins:2017jig,Jenkins:2017dyc,Liao:2020zyx}, which respects to $SU(3)_C\times U(1)_{em}$ symmetry and contains all the SM particles except the heavy $t$, $h$, $Z$ and $W$. As the same $b\to s\nu\bar{\nu}$ transition describes both $B^0\to K^{\ast0}\nu\bar{\nu}$ and $B^+\to K^+\nu\bar{\nu}$, we shall resort to solutions that can simultaneously accommodate the experimental constraints of these two processes. It is found that scenario with operator $\mathcal{O}_{L}^{\nu_\ell}$ ($\ell=e,\mu,\tau$) alone is infeasible since, they have the same dependence on the Wilson coefficient, i.e., the NP effects can be cancelled out completely, and the value of $\mathcal{B}(B^0\to K^{\ast0}\nu\bar{\nu})$ is fixed once we have known the measurement of $\mathcal{B}(B^+\to K^+\nu\bar{\nu})$---that it has been out of the current experimental upper limit. Therefore, to ensure that they have a different dependence on the NP, an additional operator $\mathcal{O}_{R}^{\nu_\ell}$ is necessary. Besides, to avoid the stringent bounds from the lepton flavor violating (LFV) processes and to reduce the number of parameters as less as possible, we also assume the lepton flavor conservation, that is, only neutrino pairs with the same flavors are considered in the low-energy Lagrangian. 

As mentioned above, we have assumed that the NP  scale $\Lambda$ (which is usually assumed to be at the level of TeV) is far higher than the EW scale $\mu_\mathrm{EW}$, physics between $\mu_\mathrm{EW}$ and $\Lambda$ can be well described by the so-called SMEFT~\cite{Buchmuller:1985jz,Grzadkowski:2010es,Brivio:2017vri}. The SMEFT contains all the SM particles and is invariant under $SU(3)_C\times SU(2)_L\times U(1)_Y$, in which the SM is treated as an effective field theory of some full theories, and the NP effects can be encoded in the higher dimension operators. Therefore, in order to connect the low-energy observables to the NP, one shall further match the LEFT operators onto the ones of the SMEFT. In this way contributions from any ultraviolet complete models can be obtained after matching onto the SMEFT and taking into account the renormalization running effects. As a byproduct of the use of the SMEFT, the same SMEFT contributing to $b\to s\nu\bar{\nu}$ will also induce LEFT operators that are relevant to processes $b\to s\ell^+\ell^-$, $b\to u_i\ell\bar{\nu}$, $u_j\to s\ell\bar{\nu}$, $u_j\to u_i\nu\bar{\nu}$, and $u_j\to u_i\ell^+\ell^-$, which implies that the constraints from $\mathcal{B}(B\to K^{(\ast)}\nu\bar{\nu})$ processes shall also subject to bounds from, e.g., $R_{K^{(\ast)}}$, $R_{D^{(\ast)}}$, $\mathcal{B}(B_s\to\tau^+\tau^-)$, $\mathcal{B}(B_{(c)}^+\to\tau^+\nu_\tau)$, $\mathcal{B}(D^+_s\to\tau^+\nu_\tau)$, $\mathcal{B}(D\to P\nu\bar{\nu})$ ($D=D^0(D^+)[D^+_s]$, $P=\pi^0(\pi^+)[K^+]$), etc.  Therefore, a combined analysis including all of the aforementioned observables is strongly called for. Note that similar studies, utilizing SMEFT to interconnect different types of $B$ decays, have been extensively discussed, for example, in Refs.~\cite{Calibbi:2015kma,Becirevic:2023aov,deGiorgi:2022vup,Allwicher:2023xba}\footnote{
We particularly refer readers to Ref.~\cite{Calibbi:2015kma}, which uses a more minimal set of SMEFT operators ($\mathcal{O}_{\ell q}^{(1)}$ and $\mathcal{O}_{\ell q}^{(3)}$) to simultaneously reconcile $b\to s\ell\ell^{(')}$, $B\to K^{(*)}\nu\overline{\nu}$, and $B\to D^{(*)}\tau\nu$ with third-generation couplings at that time. Nevertheless, as we analyze in Sec.~\ref{subsec:b2snunu}, to accommodate both the current Belle II measurement of $\mathcal{B}(B^+ \to K^+ \nu  \bar{\nu})$ and the upper limit of $\mathcal{B}(B \to K^{\ast } \nu  \bar{\nu})$, we require an additional right-handed operator $\mathcal{O}_R^{\nu_\ell}$. This necessity entails the introduction of an additional SMEFT operator, $\mathcal{O}_{\ell d}$. Therefore, the results in Ref.~\cite{Calibbi:2015kma} are refined in this work.
}. However, the compatibility of constraints from $B$ decays with those from $D$ meson decays within the same SMEFT framework has been scarcely explored in previous literature. This work aims to address this gap and provide answers to this question.

This paper is outlined as follows. In Sec.~\ref{sec:eft} we will first utilize the LEFT to describe $b\to s\nu\bar{\nu}$, and then match the LEFT operators onto the ones of SMEFT. As the same SMEFT operators can also contribute to processes $b\to s\ell^+\ell^-$, $b\to u_i\ell\bar{\nu}$, $u_j\to s\ell\bar{\nu}$, $u_j\to u_i\nu\bar{\nu}$, and $u_j\to u_i\ell^+\ell^-$, we outline the matching relations for the involved operators of LEFT and SMEFT. The theoretical expressions for the relevant observables that will be used in our final analysis are listed in Sec.~\ref{sec:phen}. Our numerical results and conclusion will be given in Sec.~\ref{sec:num} and Sec.~\ref{sec:con}, respectively.  

\section{The working frame of effective field theory}
\label{sec:eft}

In this section we will first describe the low-energy transition $b\to s\nu\bar{\nu}$ and other relevant processes like $b\to s\ell^+\ell^-$, $b\to u_i\ell\bar{\nu}$, $u_j\to s\ell\bar{\nu}$, $u_j\to u_i\nu\bar{\nu}$, and $u_j\to u_i\ell^+\ell^-$ in the LEFT, and then match the involved operators onto the ones of the SMEFT at $\mu_\mathrm{EW}$. As there are mass gaps between the low-energy scales and $\mu_\mathrm{EW}$, the renormalization evolution effects are then discussed. The Wilson coefficients can generally be separated into two parts, one from the SM and the other from NP, so throughout this paper we will use the following convention:
\begin{align}
C_i=C_{i,\mathrm{SM}}+\Delta C_i\,,
\end{align}
where $C_{i,\mathrm{SM}}$ and $\Delta C_i$ denote the SM and NP components, respectively. Throughout this work we do not concern about the charge-parity (CP) violation problem, so we will assume that all of the Wilson coefficients are real numbers.

\subsection{Description in LEFT}
 
\subsubsection{$b \to s\nu\bar{\nu}$}
\label{subsec:b2snunuLEFT}

The $B\to K^{(\ast)}\nu\bar{\nu}$ decays at the quark level are effectively described by the $b\to s\nu\bar\nu$ transition by using the $SU(3)_C\times U(1)_{em}$ invariant LEFT operators. With the assumptions that the right-handed neutrinos are absent and the neutrinos are Dirac types (so there are no scalar and tensor operators),\footnote{Coincidently, as SMEFT will link $b\to s\nu\bar{\nu}$ to $b\to s\ell^+\ell^-$, the global fits of NP in the latter in Ref.~\cite{Wen:2023pfq} also support the absence of the scalar operators.} as well as that the lepton flavor is conserved (which implies that only diagonal neutrino pairs are considered), the most general LEFT Lagrangian for $b\to s\nu\bar\nu$ at the scale $\mu_b=m_b$ reads ($\ell=e,\mu,\tau$)
\begin{align}\label{eq:b2snunu}
    \mathcal{L}_\mathrm{eff}^{b\to s\nu\bar\nu}=\frac{4G_F}{\sqrt{2}}V_{tb}V_{ts}^\ast\frac{\alpha}{4\pi}\sum_\ell\left(C^{\nu_\ell}_{L}\mathcal{O}^{\nu_\ell}_{L}+C^{\nu_\ell}_{R}\mathcal{O}^{\nu_\ell}_{R}\right)+\mathrm{h.c.}\,,
\end{align}
where
\begin{align}
\mathcal{O}^{\nu_\ell}_{L}=&(\bar{s}\gamma_\mu P_L b)(\bar{\nu}_\ell\gamma^\mu (1-\gamma_5)\nu_\ell)\,, & \mathcal{O}^{\nu_\ell}_{R}=&(\bar{s}\gamma_\mu P_R b)(\bar{\nu}_\ell\gamma^\mu (1-\gamma_5)\nu_\ell)\,.
\end{align}
In Eq.~\eqref{eq:b2snunu}, $G_F$ is the fermion coupling constant,  $\alpha=e^2/(4\pi)$ is the fine structure constant, $V_{tb}$ and $V_{ts}$ are the relevant Cabbibo-Kobyashi-Maskawa (CKM) matrix elements, and $P_{L,R}=(1\mp\gamma_5)/2$ stand for the left- and right-handed projection operators, respectively. The operator $\mathcal{O}_{R}^{\nu_\ell}$ can only arise in the presence of NP, so $C^{\nu_\ell}_{R,\mathrm{SM}}=0$, while $\mathcal{O}_{L}^{\nu_\ell}$ can originate from either the SM or NP, with $C^{\nu_\ell}_{L,\mathrm{SM}}=-6.32(7)$~\cite{Buras:2014fpa,Brod:2021hsj,Bause:2023mfe}. 

\subsubsection{Other processes}\label{subsec:otherLEFT}

As we will discuss in Sec.~\ref{subsec:SMEFT}, matching LEFT onto SMEFT will usually relate $b\to s\nu\bar{\nu}$ to other processes like $b\to s\ell^+\ell^-$, $b\to u_i\ell\bar{\nu}$, $u_j\to s\ell\bar{\nu}$, $u_j\to u_i\nu\bar{\nu}$, and $u_j\to u_i\ell^+\ell^-$, that will provide us extra available constraints and therefore enrich the content of investigation. Before proceeding discussion on the SMEFT, we will display beforehand all other relevant low-energy transitions that are interconnected through the same SMEFT operators in Eq.~\eqref{eq:SMEFT}.

As the charged lepton is a counterpart of neutrino in an $SU(2)_L$ doublet, the dileptons are always present accompanying with the dineutrinos. Hence, $[\mathcal{O}_{\ell q}^{(1)}]_{\ell\ell 23}$ and $[\mathcal{O}_{\ell q}^{(3)}]_{\ell\ell 23}$ ($[\mathcal{O}_{\ell d}]_{\ell\ell 23}$) in Eq.~\eqref{eq:SMEFT} can produce not only $\mathcal{O}_L^{\nu_\ell}$ ($\mathcal{O}_R^{\nu_\ell}$) for $b\to s\nu\bar{\nu}$, but also $\mathcal{O}_{9,10}^{\ell}$ ($\mathcal{O}_{9,10}^{\ell(\prime)}$) for $b\to s\ell^+\ell^-$, with 
\begin{align}
\mathcal{O}_9^{\ell(\prime)}=&(\bar{s}\gamma_\mu P_{L(R)}b)(\bar{\ell}\gamma^\mu\ell)\,, &
\mathcal{O}_{10}^{\ell(\prime)}=&(\bar{s}\gamma_{\mu}P_{L(R)}b)(\bar{\ell}\gamma^{\mu}\gamma_{5}\ell)\,,
\end{align}
are defined in the following low-energy effective Lagrangian
\begin{align}\label{eq:b2sll}
\mathcal{L}_\mathrm{eff}^{b\to s\ell^+\ell^-}=\frac{4G_F}{\sqrt{2}}V_{tb}V_{ts}^\ast\frac{\alpha}{4\pi}\sum_{i=9}^{10}(C_i^\ell\mathcal{O}_i^\ell+C_i^{\ell\prime}\mathcal{O}_i^{\ell\prime})+\mathrm{h.c.}\,.
\end{align}
In the SM, only $C_{9}^\ell$ and $C_{10}^\ell$ are non-zero, and the couplings to three generations of leptons are universal, which at next-to-next-to-leading logarithmic (NNLL) order are separately given by~\cite{Hou:2014dza}
\begin{align}
C_{9,\mathrm{SM}}^{\ell}\big|_{\mathrm{NNLL}}=&4.2607\,, & C_{10,\mathrm{SM}}^{\ell}\big|_{\mathrm{NNLL}}=&-4.2453\,.
\end{align}

Meanwhile, as one can see from Eq.~\eqref{eq:massbasis}, $[\mathcal{O}_{\ell q}^{(1)}]_{\ell\ell 23}$ and $[\mathcal{O}_{\ell q}^{(3)}]_{\ell\ell 23}$ can produce not only FCNC operators with down-type quarks which relate to processes like $b\to s\nu\bar{\nu}$ and $b\to s\ell^+\ell^-$, but also FCNC operators with up-type quarks which connect to processes like $c\to u\nu\bar{\nu}$ and $c\to u\ell^+\ell^-$. Note that a similar study correlating NP effects in semileptonic $\Delta C=1$ and $\Delta S=1$ processes has been published in Ref.~\cite{Fajfer:2023nmz}. The LEFT Lagrangian for the latter reads
\begin{align}\label{eq:c2u}
\mathcal{L}_\mathrm{eff}^{c\to u}=&\frac{4G_F}{\sqrt{2}}\frac{\alpha}{4\pi}\Big[C_{L}^{U\nu_\ell}(\bar{u}\gamma_\mu P_L c)(\bar{\nu}_\ell\gamma^\mu P_L\nu_\ell)+C_{L}^{U\ell}(\bar{u}\gamma_\mu P_L c)(\bar{\ell}\gamma^\mu P_L\ell)\Big]+\mathrm{h.c.}\,,
\end{align}
which describes the semileptonic decays $D\to P\nu\bar{\nu}(\ell^+\ell^-)$, as well as the purely leptonic decays $D\to\nu\bar{\nu}(\ell^+\ell^-)$. In the SM, $C_{L,\mathrm{SM}}^{U\nu_\ell}\simeq 3.0\times10^{-5}$ was calculated in Ref.~\cite{Badin:2010uh}, whereas $|C_{L,\mathrm{SM}}^{U\ell}|\lesssim0.01$ was estimated in Refs.~\cite{deBoer:2015boa,Bause:2019vpr}. Furthermore, since operator $[\mathcal{O}_{\ell q}^{(3)}]_{\ell\ell 23}$ is a triplet in the weak isospin space, that it can produce not only FCNC operators as discussed above, but also charged-current operators like $b\to u_i\ell\bar{\nu}$ and $c\to s\ell\bar{\nu}$, which provide us the investigation plenty of processes. These processes can be uniformly described by the $d_j\to u_i\ell\nu_\ell$ transition, the LEFT Lagrangian of which can be written as 
\begin{align}\label{eq:b2u}
\mathcal{L}_\mathrm{eff}^{d_j\to u_i\ell\nu_\ell}=-\frac{4G_F}{\sqrt{2}}V_{ij}(1+\Delta C_L^{ij\ell})(\bar{u}_{i}\gamma_\mu P_L d_j)(\bar{\ell}\gamma^\mu P_L\nu_\ell)+\mathrm{h.c.}\,.
\end{align}
This transition dominates the semileptonic $B$ decays like $B\to D^{(\ast)}\ell\bar{\nu}_\ell$ and charmed decays like $D\to K\ell^+\nu_\ell$, as well as the purely leptonic $B$ decays like $B_{(c)}^+\to\ell\bar{\nu}_\ell$ and charmed decays like $D_s^+\to\ell^+\nu_\ell$.

\subsection{Descriptions in SMEFT}\label{subsec:SMEFT}

From the view of LEFT, the Lagrangian in Eqs.~\eqref{eq:b2snunu}, \eqref{eq:b2sll}, \eqref{eq:c2u}, and \eqref{eq:b2u} are always valid, provided that the NP scale $\Lambda$ is well larger than the low-energy scales ($\mu=m_b$ for $B$ meson decays and $\mu=2$~GeV for charmed meson decays) but lower than $\mu_\mathrm{EW}$. In this case, these LEFT operators have no connections with each other, so we merely need to consider the bounds from individual observables. However, if we further assume that $\Lambda$ is far higher than $\mu_\mathrm{EW}$ (1~TeV for example in this work), then the EW symmetry is restored in the range between $\mu_\mathrm{EW}$ and $\Lambda$, and the operators in this range shall be of the $SU(3)_C\times SU(2)_L\times U(1)_Y$ invariant forms. If the spontaneous breaking of the EW symmetry is linearly realized, then the NP effects can be well described by the SMEFT~\cite{Buchmuller:1985jz,Grzadkowski:2010es,Brivio:2017vri}, since contributions from any ultraviolet complete model can be obtained after matching to SMEFT and taking into account the renormalization running effects. In this respect, we shall match the LEFT onto the SMEFT, which will usually interconnect the individual low-energy processes discussed in Sec.~\ref{subsec:b2snunuLEFT} and \ref{subsec:otherLEFT}.

The primary objective of this work is to address the deviations observed in $B\to K^{(\ast)}\nu\bar{\nu}$ within the framework of SMEFT. Our main focus is to identify a set of the most relevant SMEFT operators that contribute to these decays. While other flavor processes may also involve contributions from some of these operators, we specifically investigate the impacts of the selected operators on these flavor observables, while disregarding contributions from other irrelevant operators.   After a scrutiny on the full set of the SMEFT operators, we find the most relevant $SU(3)_C\times SU(2)_L\times U(1)_Y$ invariant four-fermion SMEFT operators capable of producing LEFT operators as described in Eqs.~\eqref{eq:b2snunu}-\eqref{eq:b2u} are given by\footnote{when considering up-type quark processes, the dimension-six SMEFT operator $(\bar{L}_p \gamma^\mu L_r)(\bar{u}_{Rs} \gamma_\mu u_{Rt})$ can also be matched onto the LEFT operators in Eq.~\eqref{eq:c2u} and, in principle, should be taken into account. However, the current measurements for up-type processes are either not stringent enough or almost entirely absent (see the $D \to P \nu \bar{\nu}$ processes listed in the third column of Table~\ref{tab:comparison}), providing no meaningful constraints on the Wilson coefficients in Eq.~\eqref{eq:SMEFT}. Therefore, in our numerical analysis, we predict these observables using the Wilson coefficients derived from the constraints on $\mathcal{B}(B^+ \to K^+ \nu \bar{\nu})$, $\mathcal{B}(B^0 \to K^{\ast 0} \nu \bar{\nu})$, $R_D$, $R_{D^\ast}$, and $\Delta C_9^\mu$. Similarly, if the operator $(\bar{L}_p \gamma^\mu L_r)(\bar{u}_{Rs} \gamma_\mu u_{Rt})$ were included, its corresponding Wilson coefficient would be a free parameter, leading to an indeterminate prediction. For this reason, we omit this operator from our study.}
\begin{align}\label{eq:SMEFT}
\mathcal{L}_\mathrm{SMEFT}\supset&\frac{1}{\Lambda^2}\Big[[C_{\ell d}]_{prst}(\bar{L}_p\gamma^\mu L_r)(\bar{d}_{Rs}\gamma_{\mu}d_{Rt})+[C_{\ell q}^{(1)}]_{prst}(\bar{L}_p\gamma^\mu L_r)(\bar{Q}_s\gamma_\mu Q_t)\notag\\
+&[C_{\ell q}^{(3)}]_{prst}(\bar{L}_p\gamma^\mu\tau^I L_r)(\bar{Q}_s\gamma_\mu\tau^I Q_t)\Big]\,,
\end{align}
where $\tau^I$ denote the Pauli matrices, and $L=(\nu_L,\ell_L)^T$ and $Q=(u_L,d_L)^T$ stand respectively for the left-handed lepton and quark $SU(2)_L$ doublets, while $d_R$ is the right-handed down-type quark $SU(2)_L$ singlet, with $p,r,s,t$ being the generation indices. The operators in Eq.~\eqref{eq:SMEFT} are dimension six, so the corresponding Wilson coefficients contain a suppression factor $1/\Lambda^2$, according to the power counting of SMEFT~\cite{Buchmuller:1985jz,Grzadkowski:2010es,Brivio:2017vri}. Higher dimension SMEFT operators are even more suppressed with respect to the dimension six ones, so will not be included in this work. To see more clearly the matching relations between LEFT and SMEFT, we rewrite Eq.~\eqref{eq:SMEFT} in the mass basis,
\begin{align}\label{eq:massbasis}
\mathcal{L}_\mathrm{SMEFT}\supset&\frac{1}{\Lambda^2}\Bigg\{[C_{\ell d}]_{prst}\Big[(\bar{\nu}_{Lp}\gamma^\mu\nu_{Lr})(\bar{d}_{Rs}\gamma_\mu d_{Rt})+(\bar{\ell}_{Lp}\gamma^\mu\ell_{Lr})(\bar{d}_{Rs}\gamma_\mu d_{Rt})\Big] \notag\\
+&[C_{\ell q}^{(1)}]_{prst}\Big[(\bar{\nu}_{Lp}\gamma^\mu\nu_{Lr})(\bar{d}_{Ls}\gamma_\mu d_{Lt})+(\bar{\ell}_{Lp}\gamma^\mu\ell_{Lr})(\bar{d}_{Ls}\gamma_\mu d_{Lt})\notag\\
+&V_{is}V_{jt}^\ast\big[(\bar{\nu}_{Lp}\gamma^\mu\nu_{Lr})(\bar{u}_{Li}\gamma_\mu u_{Lj})+(\bar{\ell}_{Lp}\gamma^\mu\ell_{Lr})(\bar{u}_{Li}\gamma_\mu u_{Lj})\big]\Big]\notag\\
+&[C_{\ell q}^{(3)}]_{prst}\Big[-(\bar{\nu}_{Lp}\gamma^\mu\nu_{Lr})(\bar{d}_{Ls}\gamma_\mu d_{Lt})+(\bar{\ell}_{Lp}\gamma^\mu\ell_{Lr})(\bar{d}_{Ls}\gamma_\mu d_{Lt})\notag\\
+&2V_{is}(\bar{\ell}_{Lp}\gamma^\mu\nu_{Lr})(\bar{u}_{Li}\gamma_\mu d_{Lt})+2V_{jt}^\ast(\bar{\nu}_{Lp}\gamma^\mu\ell_{Lr})(\bar{d}_{Ls}\gamma_\mu u_{Lj})\notag\\
+&V_{is}V_{jt}^\ast\big((\bar{\nu}_{Lp}\gamma^\mu\nu_{Lr})(\bar{u}_{Li}\gamma_\mu u_{Lj})-(\bar{\ell}_{Lp}\gamma^\mu \ell_{Lr})(\bar{u}_{Li}\gamma_\mu u_{Lj})\big)\Big]\Bigg\}\,,
\end{align}
where we have chosen to work under the down basis, in which both the down quark and the charged lepton Yukawa couplings are diagonal, while the right-handed fermions are in the mass basis. For terms involved CKM matrix elements $V_{is(jt)}$, a sum over the up-type quark indices $i,j=u,c,t$ is understood. As we beforehand discussed in Sec.~\ref{subsec:otherLEFT}, the SMEFT operators in Eq.~\eqref{eq:massbasis} can produce not only LEFT operators contributing to $b\to s\nu\bar{\nu}$, but also extra operators that contribute to processes like $b\to s\ell^+\ell^-$, $b\to u_i\ell\bar{\nu}$, $u_j\to s\ell\bar{\nu}$, $u_j\to u_i\nu\bar{\nu}$, and $u_j\to u_i\ell^+\ell^-$, which will connect $B\to K^{(\ast)}\nu\bar{\nu}$ to $B\to K^{(\ast)}\ell^+\ell^-$, $B\to D^{(\ast)}\ell\nu_\ell$, $B_s\to\ell^+\ell^-$, $B_{(c)}^+\to\ell^+\nu_\ell$, $D^+_s\to\ell^+\nu_\ell$, $D\to P\nu\bar{\nu}$, etc. Therefore, a combined analysis including these processes are necessary. To this end, one shall first perform the match the LEFT operators onto the ones of the SMEFT.

\subsection{Matching LEFT onto SMEFT}

Generally, since there are mass gaps between the low-energy scales  and $\mu_\mathrm{EW}$, we shall take the renormalization running effects into account, which are governed by the following renormalization group equations (RGEs)~\cite{Jenkins:2017dyc}
\begin{align}
\frac{d C_{L(R)}^{\nu_\ell}}{d\log\mu}=&0\,, & \frac{d C_{L(R)}^{D\ell}}{d\log\mu}=&(\frac{4}{3}q_e^2+(-)12q_dq_e)\frac{\alpha}{4\pi}C_{L(R)}^{D\ell}\,, & \frac{d C_{L}^{ij\ell}}{d\log\mu}=&6q_uq_e\frac{\alpha}{4\pi}C_{L}^{ij\ell}\,, \notag\\
\frac{d C_{L}^{U\nu_\ell}}{d\log\mu}=&0\,, &
\frac{d C_{L}^{U\ell}}{d\log\mu}=&(\frac{4}{3}q_e^2+12q_uq_e)\frac{\alpha}{4\pi}C_{L}^{U\ell}\,, &&
\end{align}
where the charge numbers $q_{e,u,d}=-1\,,+2/3\,,-1/3$, respectively. Note that for Wilson coefficients in the second RGE, we have used the linear combinations $C_{L(R)}^{D\ell}=C_9^{\ell(\prime)}-C_{10}^{\ell(\prime)}$ instead of the individual $C_{9}^{\ell(\prime)}$ and $C_{10}^{\ell(\prime)}$. Here only one-loop anomalous dimensions are considered in the RGEs, and we do not consider the mixing effects from other irrelevant operators. It is clear that the anomalous dimension is zero for $C_{L(R)}^{\nu_\ell}$ and $C_{L}^{U\nu_\ell}$, so there have no running effects, while only QED anomalous dimensions are involved in the rest of coefficients, which implies that the running effects are very small and can be neglected safely. With this observation we straightforwardly obtain the following matching relations:
\begin{align}\label{eq:match}
\frac{8G_F}{\sqrt{2}}V_{tb}V_{ts}^\ast\frac{\alpha}{4\pi}\Delta C_{L}^{\nu_\ell}=&\frac{[C_{\ell q}^{(1)}]_{\ell\ell 23}-[C_{\ell q}^{(3)}]_{\ell\ell 23}}{\Lambda^2}\,,\notag\\
\frac{8G_F}{\sqrt{2}}V_{tb}V_{ts}^\ast\frac{\alpha}{4\pi}\Delta C_{R}^{\nu_\ell}=&\frac{[C_{\ell d}]_{\ell\ell 23}}{\Lambda^2} \,,\notag\\
\frac{4G_F}{\sqrt{2}}V_{tb}V_{ts}^\ast\frac{\alpha}{4\pi} (\Delta C_9^\ell-\Delta C_{10}^\ell)=&\frac{[C_{\ell q}^{(1)}]_{\ell\ell 23}+[C_{\ell q}^{(3)}]_{\ell\ell 23}}{\Lambda^2}\,,\notag\\
\frac{4G_F}{\sqrt{2}}V_{tb}V_{ts}^\ast\frac{\alpha}{4\pi} (\Delta C_9^{\ell\prime}-\Delta C_{10}^{\ell\prime})=&\frac{[C_{\ell d}]_{\ell\ell 23}}{\Lambda^2}\,,\notag\\
-\frac{4G_F}{\sqrt{2}}\frac{V_{ib}^\ast}{2V_{is}}\Delta C_{L}^{ib\ell\ast}=&\frac{[C_{\ell q}^{(3)}]_{\ell\ell 23}}{\Lambda^2}\,, \notag\\
-\frac{4G_F}{\sqrt{2}}\frac{V_{js}}{2V_{jb}^\ast}\Delta C_{L}^{js\ell}=&\frac{[C_{\ell q}^{(3)}]_{\ell\ell 23}}{\Lambda^2}\,,\notag\\
\frac{4G_F}{\sqrt{2}}\frac{\alpha}{4\pi}\frac{1}{V_{is}V_{jb}^\ast} \Delta C_{L}^{U\nu_\ell}=&\frac{[C_{\ell q}^{(1)}]_{\ell\ell 23}+[C_{\ell q}^{(3)}]_{\ell\ell 23}}{\Lambda^2}\,, \notag\\
\frac{4G_F}{\sqrt{2}}\frac{\alpha}{4\pi}\frac{1}{V_{is}V_{jb}^\ast} \Delta C_{L}^{U\ell}=&\frac{[C_{\ell q}^{(1)}]_{\ell\ell 23}-[C_{\ell q}^{(3)}]_{\ell\ell 23}}{\Lambda^2}\,.
\end{align} 
It is clear from Eq.~\eqref{eq:match} that,  $\Delta C_L^{\nu_\ell}$ ($\Delta C_L^{D\ell}$) share the same SMEFT Wilson coefficients with $\Delta C_L^{U\ell}$ ($\Delta C_L^{U\nu_\ell}$), similar observations also hold between $\Delta C_R^{\nu_\ell}$ and $\Delta C_R^{D\ell}$ as well as $\Delta C_L^{ib\ell}$ and $\Delta C_L^{js\ell}$, up to different constants from matchings. These relations can be further refined after taking into account some of the most established phenomenological constraints, e.g., from the $b\to s\ell^+\ell^-$ processes, $R_{D^{(\ast)}}$, etc., more details will be discussed in next section.

\subsection{Comments on other SMEFT operators}

Before finishing this section we make some comments on the following  dimension-six SMEFT operators with quark-bilinears and the Higgs $H$,
\begin{align}\label{eq:Hq}
\mathcal{L}_\mathrm{SMEFT}\supset& \frac{1}{\Lambda^2}\Big[[C_{H d}]_{pr}(\bar{d}_{Rp}\gamma_{\mu}d_{Rr})(H^\dagger i\overleftrightarrow{D}^\mu H)+
[C_{H q}^{(1)}]_{pr}(\bar{Q}_p\gamma_\mu Q_r)(H^\dagger i\overleftrightarrow{D}^\mu H)\notag\\
+&[C_{H q}^{(3)}]_{pr}(\bar{Q}_p\gamma_\mu\tau^I Q_r)(H^\dagger i\overleftrightarrow{D}^\mu\tau^I H)\Big]\,,
\end{align}
where $H^\dagger i\overleftrightarrow{D}^\mu(\tau^I)H\equiv H^\dagger(iD^\mu(\tau^I)H)-(iD^\mu H^\dagger)(\tau^I)H$. These operators can contribute to the electroweak and top observables, as well as the neutral meson mixings (via double insertion of the operators), so the corresponding Wilson coefficients shall subject to the stringent constraints from these observables, the relevant analysis see, e.g., Refs.~\cite{Falkowski:2014tna,Ethier:2021bye,Garosi:2023yxg}. Clearly, these operators can also have impacts on the LEFT operators listed in Eqs.~\eqref{eq:b2snunu}, \eqref{eq:b2sll}, \eqref{eq:c2u}, and \eqref{eq:b2u}, through correcting the effective couplings of $Z$ and $W$ to the quark currents, as discussed in Ref.~\cite{Allwicher:2023xba}. However, the couplings of $Z$ and $W$ to lepton currents remain universal and are not affected by these operators, so after integrating out the heavy gauge bosons, the total Wilson coefficients will be universal, too. Since the SMEFT operators in Eq.~\eqref{eq:Hq} also link $b\to s\nu\bar{\nu}$ to $b\to s\ell^+\ell^-$, the global fits of the latter in Ref.~\cite{Wen:2023pfq} show that $C_{9,10}^{\ell\prime}=0$ for $\ell=e,\mu$, with the matchings in Eq.~\eqref{eq:match} one immediately obtains that $C_R^{\nu_\ell}=0$ (so $C_R^{\nu_\tau}=0$ as well, if couplings are universal), too.\footnote{
In this work, we have explicitly set $C_{9,10}^{\ell\prime}$ to zero, and by matching to SMEFT operators, this subsequently implies that $C_R^{\nu_\ell} = 0$. If we relax this condition and allow $C_{9,10}^{\ell\prime}$ to vary within their allowed ranges, an enhancement in $\mathcal{B}(B^+ \to K^+ \nu \bar{\nu})$ of up to $20\%$ could be obtained using the constraints from $\mathcal{B}(B_s\to\mu^+\mu^-)$ and $\Delta m_{B_s}$. Such an enhancement is obviously insufficient to accommodate the Belle-II excess, detailed analysis see Ref.~\cite{Allwicher:2023xba}.} However, this is in conflict with that a nonzero $C_R^{\nu_\ell}$ is necessary to simultaneously accommodate both the current bounds on $\mathcal{B}(B^+\to K^{+}\nu\bar{\nu})$ and $\mathcal{B}(B\to K^{\ast}\nu\bar{\nu})$, as we will discuss in Sec.~\ref{subsec:b2snunu}. Similarly, the requirement of lepton flavor universality (LFU) violating interactions in $R_{D^{(\ast)}}$ also excludes such an infeasible scenario with the universal couplings. Therefore, operators listed in Eq.~\eqref{eq:Hq} will not be included in our analysis.

\section{Observables}
\label{sec:phen}

In this section we will detailedly show all the formulas, which are expressed in terms of the LEFT Wilson coefficients, for observables involved in this paper. We do not discuss dineutrino decays like $B_s\to\nu\bar{\nu}$ and $D^0\to\nu\bar{\nu}$ since, on the one hand, the lack of measurements, and on the other hand, contributions from vector operators to the branching ratios are helicity suppressed by two powers of the small neutrino mass, and negligible~\cite{Bause:2020xzj}. 

\subsection{$B\to K^{(\ast)}\nu\bar{\nu}$}\label{subsec:b2snunu}

With the LEFT Lagrangian~\eqref{eq:b2snunu} at hand, the calculation of branching ratio of $B\to K^{(\ast)}\nu\bar{\nu}$ decays is straightforward. To this end, we shall first provide the differential branching ratios. One has, for $B^+\to K^+\nu\bar{\nu}$, 
\begin{align}\label{eq:B2Knunu}
\frac{d\mathcal{B}(B^+\to K^+\nu\bar{\nu})}{dq^2}=\sum_\ell\frac{\tau_{B^+}G_F^2\alpha^2}{768\pi^5m_B^3}|V_{tb}V_{ts}^\ast|^2\lambda^{\frac{3}{2}}(q^2,m_B^2,m_K^2)\left[f_+(q^2)\right]^2\left|C_L^{\nu_\ell}+C_R^{\nu_\ell}\right|^2\,,
\end{align}
and for $B^0\to K^{\ast0}\nu\bar{\nu}$,
\begin{align}\label{eq:B2K*nunu}
\frac{d\mathcal{B}(B^0\to K^{\ast0}\nu\bar{\nu})}{dq^2}=&\sum_\ell\frac{\tau_{B^0}G_F^2\alpha^2}{384\pi^5m_B^3}|V_{tb}V_{ts}^\ast|^2\lambda^{\frac{1}{2}}(q^2,m_B^2,m_{K^\ast}^2)(m_B+m_{K^\ast})^2q^2 \notag\\
&\times\Bigg\{\left(\left[A_1(q^2)\right]^2+\frac{32m_B^2m_{K^\ast}^2}{q^2(m_B+m_{K^\ast})^2}\left[A_{12}(q^2)\right]^2\right)\left|C_L^{\nu_\ell}-C_R^{\nu_\ell}\right|^2 \notag\\
&~~~~+\frac{\lambda(q^2,m_B^2,m_{K^\ast}^2)}{(m_B+m_{K^\ast})^4}\left[V(q^2)\right]^2\left|C_L^{\nu_\ell}+C_R^{\nu_\ell}\right|^2\Bigg\}\,,
\end{align}
where $0<q^2<(m_B-m_{K^{(\ast)}})^2$ is the invariant mass of the neutrino pair, $\lambda(a,b,c)=a^2+b^2+c^2-2(ab+ac+bc)$ is the usual K{\"a}ll{\'e}n function. In Eqs.~\eqref{eq:B2Knunu} and~\eqref{eq:B2K*nunu}, one needs to sum over the couplings to three generation neutrinos, which in the SM are universal that the predictions can be easily obtained by just inputting the SM coefficient $C^{\nu_\ell}_{L,\mathrm{SM}}$ and replacing the summation by a factor $3$. The explicit expressions of the $B^+\to K^+$ vector form factor $f_+(q^2)$, as well as the $B^0\to K^{\ast0}$ form factors $A_1(q^2)$, $A_{12}(q^2)$ and $V(q^2)$ are adopted from Ref.~\cite{Buras:2014fpa} (in the numerical analysis we will use the recent updated results in Ref.~\cite{Becirevic:2023aov}).

After integrating over $q^2$ in the differential decay rates Eqs.~\eqref{eq:B2Knunu} and~\eqref{eq:B2K*nunu}, we obtain the following numerical expressions:
\begin{align}
\mathcal{B}(B^+\to K^+\nu\bar{\nu})=&3.46\times10^{-8}\sum_\ell\left|C^{\nu_\ell}_{L}+C^{\nu_\ell}_{R}\right|^2  \,,\\
\mathcal{B}(B^0\to K^\ast\nu\bar{\nu})=&6.84\times10^{-8}\sum_\ell\left|C^{\nu_\ell}_{L}-C^{\nu_\ell}_{R}\right|^2+1.36\times10^{-8}\sum_\ell\left|C^{\nu_\ell}_{L}+C^{\nu_\ell}_{R}\right|^2\,.
\end{align}
Using the input $C^{\nu_\ell}_{L,\mathrm{SM}}=-6.32(7)$, the SM predictions yield $\mathcal{B}(B^+\to K^+\nu\bar{\nu})_\mathrm{SM}=(4.43\pm0.31)\times10^{-6}$ and $\mathcal{B}(B^0\to K^\ast\nu\bar{\nu})_\mathrm{SM}=(9.47\pm1.40)\times10^{-6}$, respectively~\cite{Becirevic:2023aov}. It is clear from above equations that scenario with NP appears only in operator $\mathcal{O}^{\nu_\ell}_{L}$ is not sufficient to explain the current upper constraint of $\mathcal{B}(B^0\to K^{\ast0}\nu\bar{\nu})$, which can be seen as the coefficient $C_L^{\nu_\ell}$ is completely cancelled out in the ratio, 
\begin{align}\label{eq:ratio}
R=\frac{8.20\times10^{-8}\sum\limits_{\ell}\left|C_L^{\nu_\ell}\right|^2}{3.46\times10^{-8}\sum\limits_{\ell}\left|C_L^{\nu_\ell}\right|^2}\simeq2.37\,,
\end{align}
where we have set $C_R^{\nu_\ell}=0$. Using the Belle II measurement $\mathcal{B}(B^+\to K^+\nu\bar{\nu})_\mathrm{exp}=(2.3\pm0.7)\times 10^{-5}$~\cite{Belle-II:2023esi}, one can immediately deduce from Eq.~\eqref{eq:ratio} that $\mathcal{B}(B^0\to K^{\ast0}\nu\bar{\nu})=(5.5\pm1.7)\times 10^{-5}$, which has been out of the upper limit of the current experiment bound $\mathcal{B}(B^0\to K^{\ast0}\nu\bar{\nu})_\mathrm{exp}<1.8\times 10^{-5}$~\cite{Belle:2017oht}. Therefore, to reconcile such an issue so that the two decay rates have a different dependence on the NP parameters, a second operator $\mathcal{O}^{\nu_\ell}_{R}$ has to be included.

As the $K^{\ast0}$ longitudinal polarization fraction $F_L$ in ${\cal B}(B^0\to K^{\ast0}\nu\bar\nu)$ decay can give very complementary information besides the branching ratio, we will also give a prediction to this observable, which can be compared to the predictions from the previous literatures, e.g., Refs~\cite{Altmannshofer:2009ma,Buras:2014fpa}. Here, we are interested in the integrated form of $F_L(B^0\to K^{\ast0}\nu\bar\nu)$, which is defined as:
\begin{align}
\langle F_L(B^0\to K^{\ast0}\nu\bar\nu)\rangle=\frac{\int_0^{(m_B-m_{K^{\ast}})^2}\frac{d\Gamma_L(B^0\to K^{\ast0}\nu\bar\nu)}{dq^2}dq^2}{\int_0^{(m_B-m_{K^{\ast}})^2}\frac{d\Gamma(B^0\to K^{\ast0}\nu\bar\nu)}{dq^2}dq^2}
\end{align}
where
\begin{align}
\frac{d\Gamma_L(B^0\to K^{\ast0}\nu\bar\nu)}{dq^2}=&\sum_\ell\frac{\tau_{B^0}G_F^2\alpha^2}{384\pi^5m_B^3}|V_{tb}V_{ts}^\ast|^2\lambda^{\frac{1}{2}}(q^2,m_B^2,m_{K^\ast}^2)(m_B+m_{K^\ast})^2q^2 \notag\\
&\times\frac{32m_B^2m_{K^\ast}^2}{q^2(m_B+m_{K^\ast})^2}\left[A_{12}(q^2)\right]^2\left|C_L^{\nu_\ell}-C_R^{\nu_\ell}\right|^2\,,\\
\frac{d\Gamma(B^0\to K^{\ast0}\nu\bar\nu)}{dq^2}=&\frac{d\mathcal{B}(B^0\to K^{\ast0}\nu\bar{\nu})}{\tau_{B^0}\,dq^2}\,.
\end{align}
The numerical prediction of $\langle F_L(B^0\to K^{\ast0}\nu\bar\nu)\rangle$ will be presented in Table~\ref{tab:comparison}. The comparison between this value to the SM predictions obtained from Refs~\cite{Altmannshofer:2009ma,Buras:2014fpa} is shown in Figure~\ref{fig:reprediction}.

\subsection{$B\to D^{(\ast)}\ell\bar{\nu}_\ell$}

The charged-current processes $\mathcal{B}(B\to D^{(\ast)}\ell\bar{\nu}_\ell)$ are usually employed to measure the LFU with the following definitions:
\begin{align}
R_{D^{(\ast)}}\equiv\frac{\mathcal{B}(B\to D^{(\ast)}\tau\bar{\nu}_\tau)}{\mathcal{B}(B\to D^{(\ast)}\ell\bar{\nu}_\ell)}\,,
\end{align}
where in the denominator, $\ell=e,\mu$. The $R_{D^{(\ast)}}$ have been measured by BaBar~\cite{BaBar:2012obs,BaBar:2013mob}, Belle~\cite{Belle:2015qfa,Belle:2016dyj,Belle:2017ilt,Belle:2019gij,Belle:2019rba}, and LHCb~\cite{LHCb:2015gmp,LHCb:2017smo,LHCb:2017rln}, and the results show that all of the measurements are always excess compared with the SM predictions. The world average values of the measurements are given by HFLAV group~\cite{HFLAV:2022esi}, with
\begin{align}
R_D^\mathrm{ave}=&0.344\pm0.026\,, & R_{D^\ast}^\mathrm{ave}=&0.285\pm0.012\,,
\end{align}
and the $R_D-R_{D^\ast}$ correlation of $-0.39$, the combination of which deviates from the SM prediction~\cite{HFLAV:2022esi}
\begin{align}
R_D^\mathrm{SM}=&0.298\pm0.004\,, & R_{D^\ast}^\mathrm{SM}=&0.254\pm0.005\,,
\end{align}
at a level of $3.17\sigma$. The NP effects are encoded in the differential decay widths of $\mathcal{B}(B\to D^{(\ast)}\ell\bar{\nu}_\ell)$ decays,
\begin{align}
\frac{d\Gamma(B\to D\ell\bar{\nu}_\ell)}{dq^2}=&\frac{G_F^2|V_{cb}|^2}{192\pi^3m_B^3}q^2\lambda^{\frac{1}{2}}(q^2,m_B^2,m_D^2)\left(1-\frac{m_\ell^2}{q^2}\right)^2\left|1+\Delta C_{L}^{cb\ell}\right|^2 \notag\\
\times&\left[\left(1+\frac{m_\ell^2}{2q^2}\right)H_{V,0}^{s\,2}+\frac{3}{2}\frac{m_\ell^2}{q^2}H_{V,t}^{s\,2}\right]\,,\label{eq:RD}\\ 
\frac{d\Gamma(B\to D^\ast\ell\bar{\nu}_\ell)}{dq^2}=&\frac{G_F^2|V_{cb}|^2}{192\pi^3m_B^3}q^2\lambda^{\frac{1}{2}}(q^2,m_B^2,m_{D^\ast}^2)\left(1-\frac{m_\ell^2}{q^2}\right)^2\left|1+\Delta C_{L}^{cb\ell}\right|^2\notag\\
\times&\left[\left(1+\frac{m_\tau^2}{2q^2}\right)(H_{V,+}^2+H_{V,-}^2+H_{V,0}^2)+\frac{3m_\tau^2}{2q^2}H_{V,t}^2\right]\,,\label{eq:RDast}
\end{align}
where $H_{V,i}^{s}$ ($i=0,t$) and $H_{V,j}$ ($j=\pm,0,t$) stand for the hadronic helicity amplitudes of $B\to D$ and $B\to D^\ast$ transitions, separately, the explicit expressions of which can be found in Refs.~\cite{Tanaka:2012nw,Sakaki:2013bfa}. Here we adopt a usual assumption which is accepted by most of the literatures that NP contributions present only in the third generation of leptons, that is, $\Delta C_L^{cb\tau}\neq0$ and $\Delta C_L^{cbe}=\Delta C_L^{cb\mu}=0$, which corresponds to $[C_{\ell q}^{(3)}]_{3323}\neq0$ and $[C_{\ell q}^{(3)}]_{1123}=[C_{\ell q}^{(3)}]_{2223}=0$. As discussed in Ref.~\cite{Murgui:2019czp}, such a consensus is motivated by the absence of significant discrepancy between the measurements and the SM predictions on processes involving light leptons, e.g., $b\to c\ell\bar{\nu}_\ell$ ($\ell=e,\mu$) decays~\cite{Jung:2018lfu}. Integrating over $q^2$ in Eqs.~\eqref{eq:RD} and \eqref{eq:RDast} yields the following numerical expressions~\cite{Iguro:2022yzr}:
\begin{align}
R_D=&R_D^\mathrm{SM}\left|1+\Delta C_L^{cb\tau}\right|^2\,, & 
R_{D^\ast}=&R_{D^\ast}^\mathrm{SM}\left|1+\Delta C_L^{cb\tau}\right|^2\,.
\end{align}
Obviously, this will be a feasible scenario with a single operator $\mathcal{O}_L^{cb\tau}$, since the following ratios,
\begin{align}
\frac{R_D^\mathrm{ave}}{R_D^\mathrm{SM}}=&1.15\pm0.09\,, & \frac{R_{D^\ast}^\mathrm{ave}}{R_{D^\ast}^\mathrm{SM}}=&1.12\pm0.06\,,
\end{align}
are very closed, which implies that the discrepancy can be naturally accommodated by a common Wilson coefficient $\Delta C_L^{cb\tau}$.

\subsection{$B\to K^{(\ast)}\ell^+\ell^-$}
\label{subsec:C9}

In the rare $B$ decays, the LFU can be tested not only in $R_{D^{(\ast)}}$, which are governed by the charged-current operators, but also in $R_{K^{(\ast)}}$, which are defined as
\begin{align}
R_{K^{(\ast)}}\equiv\frac{\mathcal{B}(B\to K^{(\ast)}\mu^+\mu^-)}{\mathcal{B}(B\to K^{(\ast)}e^+e^-)}\,,
\end{align}
in which the semileptonic decays $B\to K^{(\ast)}\ell^+\ell^-$ are effectively described by the FCNC processes $b\to s\ell^+\ell^-$. The explicit expressions for $\mathcal{B}(B\to K^{(\ast)}\ell^+\ell^-)$ are can be found in Refs.~\cite{Bobeth:2007dw,Altmannshofer:2008dz}, which are also collected in the appendix of Ref.~\cite{Wen:2023pfq}. The deviations between the measurements and the SM predictions on $R_{K^{(\ast)}}$ are long-standing~\cite{LHCb:2014vgu,LHCb:2017avl,BELLE:2019xld,LHCb:2019hip,LHCb:2021lvy}, until the state-of-the-art results on this respect reported by LHCb in 2022 indicating that the well-known anomalies in $R_{K^{(\ast)}}$ have faded away~\cite{LHCb:2022vje}. It seems that the possibility for NP existing in $b\to s\ell^+\ell^-$ processes has been ruled out. However, Using around 200 observables in leptonic and semileptonic decays of $B$ mesons and bottom baryons, including $B_{s,d}\to\ell^+\ell^-$, $B\to K^{(\ast)}\ell^+\ell^-$, $B\to X_s\ell^+\ell^-$, $\Lambda_b\to\Lambda\ell^+\ell^-$, two of us in Ref.~\cite{Wen:2023pfq} show that there still leaves room for NP present in $\Delta C_9^{\ell}$ ($\ell=e,\mu$), which is deviated from the SM at the level of around or more than $4\sigma$. For simplicity, in this work we will quote the lepton-universal scenario (S-II) in Ref.~\cite{Wen:2023pfq}, in which the degrees of freedom are $C_{9,10,S,P}^{\mu(\prime)}$ and $C_{7,8}^{(\prime)}$ with $C_{9,10,S,P}^{\mu(\prime)}=C_{9,10,S,P}^{e(\prime)}$, where $C_{7}^{(')}$, $C_{8}^{(')}$, $C_{S}^{\mu(')}$, and $C_{P}^{\mu(')}$ denote the Wilson coefficients of the electromagnetic-dipole, chromomagnetic-dipole, scalar, and pseudo-scalar operators, respectively. The global fit of S-II yielded
\begin{align}
\Delta C_9^{e}=\Delta C_9^{\mu}=-0.789_{-0.210}^{+0.198}\,,
\end{align}
while the remaining Wilson coefficients are consistent with zero. Comparing these results with the matching relations in Eq.~\eqref{eq:match}, one can immediately deduce that $[C_{\ell q}^{(1)}]_{1123}=[C_{\ell q}^{(1)}]_{2223}\neq0$ and $[C_{\ell d}]_{1123}=[C_{\ell d}]_{2223}=0$, the latter implies that $C_R^{\nu_e}=C_R^{\nu_\mu}=0$. This, together with the requirement that a nonzero $C_R^{\nu_\ell}$ must be included to simultaneously explain the experimental bounds on $\mathcal{B}(B\to K^{(\ast)}\nu\bar{\nu})$, renders that only a nonvanishing $C_R^{\nu_\tau}$ (or equivalently, $[C_{\ell d}]_{3323}\neq0$) can fulfil all of these conditions. Our opinion is consistent with that of Ref.~\cite{Bause:2023mfe}.

\subsection{$B_{q}^+\to\tau^+\nu_\tau$ and $B_s\to\tau^+\tau^-$}

The charged- and neutral-current operators discussed above also affect the leptonic $B$ decays, so we can include them in our analysis as a complementarity. 

For the charged-current process, since only $[C_{\ell q}^{(3)}]_{3323}$ exists, we merely consider $B$ mesons decay into the third generation leptons. From Eq.~\eqref{eq:SMEFT} one can see that the leptonic processes sharing the same SMEFT operator $[\mathcal{O}_{\ell q}^{(3)}]_{3323}$ with the semileptonic processes $B\to D^{(\ast)}\tau\bar{\nu}_\tau$ are $B_{(c)}^+\to\tau^+\nu_\tau$ decays, the branching ratio of which can be uniformly written as
\begin{align}
\mathcal{B}(B_{q}^+\to\tau^+\nu_\tau)=\tau_{B_{q}^+}\frac{G_F^2}{8\pi}|V_{qb}|^2f_{B_{q}^+}^2 m_\tau^2 m_{B_{q}^+}\left(1-\frac{m_\tau^2}{m_{B_{q}^+}^2}\right)^2\left|1+\Delta C_L^{qb\tau}\right|^2\,,
\end{align}
with $q=u$ for $B^+\to\tau^+\nu_\tau$ and $q=c$ for $B_c^+\to\tau^+\nu_\tau$ decays, respectively. For branching ratio of $B^+\to\tau^+\nu_\tau$, the average of the measurements from BaBar~\cite{BaBar:2009wmt,BaBar:2012nus} and Belle~\cite{Belle:2012egh,Belle:2015odw} gives $\mathcal{B}(B^+\to\tau^+\nu_\tau)_\mathrm{exp}=(1.09\pm0.24)\times10^{-4}$~\cite{ParticleDataGroup:2022pth}, which is consistent with the SM prediction $\mathcal{B}(B^+\to\tau^+\nu_\tau)_\mathrm{SM}=(0.87\pm0.05)\times10^{-4}$~\cite{Fedele:2023gyi}, obtained by using the $N_f=2+1+1$ lattice result of decay constant $f_{B^+}=190.0\pm1.3~\mathrm{MeV}$ from the Flavour Lattice Averaging Group (FLAG)~\cite{FlavourLatticeAveragingGroupFLAG:2021npn}, and $|V_{ub}|=(3.70\pm0.11)\times10^{-3}$ from the UTfit collaboration~\cite{UTfit:2022hsi}. Similarly, for $B_c^+\to\tau^+\nu_\tau$ decay, by using $f_{B_c^+}=427\pm6~\mathrm{MeV}$ from the lattice~\cite{McNeile:2012qf,Colquhoun:2015oha}, and $|V_{cb}|= (42.22\pm0.51)\times10^{-3}$ from the UTfit collaboration~\cite{UTfit:2022hsi}, the SM prediction is $\mathcal{B}(B_c^+\to\tau^+\nu_\tau)_\mathrm{SM}=(2.29\pm0.09)\times10^{-2}$~\cite{Fedele:2023gyi}. However, the measurement on such a respect is still missing. Therefore we will make a prediction on this observable with input $[C_{\ell q}^{(3)}]_{3323}$ obtained from previous constraints. Our prediction may be tested in the Tera-Z machines of the future experiments, e.g., from the CEPC~\cite{Zheng:2020ult} and FCC-ee~\cite{Amhis:2021cfy,Fedele:2023gyi}.

As for FCNC process $b\to s\ell^+\ell^-$, the decays of $B_{s}$ meson into the first and second generation of leptons have been included in Sec.~\ref{subsec:C9}, now let us focus on the leptonic $B_s$ meson decay into the third generation of leptons, i.e., $B_s\to\tau^+\tau^-$. The branching ratio of $B_s\to\tau^+\tau^-$ reads
\begin{align}
\mathcal{B}(B_s\to\tau^+\tau^-)=\tau_{B_s}\frac{G_F^2\alpha^2}{16\pi^3}|V_{ts}V_{tb}^\ast|^2f_{B_s}^2m_{B_s}\sqrt{1-\frac{4m_\tau^2}{m_{B_s}^2}}m_\tau^2|C_{10}^\tau-C_{10}^{\tau\prime}|^2\,,
\end{align}
where $C_{10}^{\tau}$ and $C_{10}^{\tau\prime}$ relate to the combination of $[C_{\ell q}^{(1)}]_{3323}$ and $[C_{\ell q}^{(3)}]_{3323}$ and $[C_{\ell d}]_{3323}$, respectively. The coefficients $C_{9}^{\tau}$ and $C_{9}^{\tau\prime}$ do not participate in the decay due to the parity conservation of the strong interaction. To further reduce the degrees of freedom, we will assume that the irrelevant parameters couplings $\Delta C_{9}^{\tau(\prime)}$ fulfil the following universality condition:
\begin{align}
\Delta C_{9}^{e}=&\Delta C_{9}^{\mu}=\Delta C_{9}^{\tau}=-0.789_{-0.210}^{+0.198}\,, & \Delta C_{9}^{e\prime}=&\Delta C_{9}^{\mu\prime}=\Delta C_{9}^{\tau\prime}=0\,.
\end{align}
To date, only an upper limit $\mathcal{B}(B_s\to\tau^+\tau^-)_\mathrm{exp}<6.8\times10^{-3}$ at $95\%$ CL is given by LHCb~\cite{LHCb:2017myy}, which is four orders of magnitude larger than the SM prediction $\mathcal{B}(B_s\to\tau^+\tau^-)_\mathrm{SM}=(7.73\pm0.49)\times10^{-7}$~\cite{Bobeth:2013uxa}, leaving therefore a large room for NP. The future LHCb hopes to improve the current sensitivity to about $5\times10^{-4}$~\cite{Albrecht:2017odf}, whereas CEPC may even reach a sensitivity with order of $10^{-5}$~\cite{CEPCStudyGroup:2018ghi}. Our prediction on this observable with entries from the previous constraints may be tested in these future colliders.

\subsection{$D_{s}^+\to\tau^+\nu_\tau$ and $D\to P\nu\bar{\nu}$}

The same SMEFT operators in Eq.~\eqref{eq:SMEFT} can also lead to leptonic and semileptonic charmed meson decays. For leptonic decays, we focus on the charged-current process involving the third generation leptons, i.e., $D_{s}^+\to\tau^+\nu_\tau$. The branching ratio of $D_{s}^+\to\tau^+\nu_\tau$ is
\begin{align}
\mathcal{B}(D_{s}^+\to\tau^+\nu_\tau)=\tau_{D_{s}^+}\frac{G_F^2}{8\pi}|V_{cs}|^2f_{D_{s}^+}^2 m_\tau^2 m_{D_{s}^+}\left(1-\frac{m_\tau^2}{m_{D_{s}^+}^2}\right)^2\left|1+\Delta C_L^{cs\tau}\right|^2\,,
\end{align}
in which the coefficient $\Delta C_L^{cs\tau}$ is related to $[C_{\ell q}^{(3)}]_{3323}$. The latest measurement on this branching ratio is given by BESIII collaboration, with $\mathcal{B}(D_{s}^+\to\tau^+\nu_\tau)_\mathrm{BESIII}=(5.37\pm0.17_\mathrm{stat}\pm0.15_\mathrm{syst})\%$~\cite{BESIII:2023ukh}, which is consistent with the world average given by the Particle Data Group (PDG), $\mathcal{B}(D_{s}^+\to\tau^+\nu_\tau)_\mathrm{ave}=(5.32\pm0.11)\%$~\cite{ParticleDataGroup:2022pth}, as well as the the SM prediction $\mathcal{B}(D_{s}^+\to\tau^+\nu_\tau)_\mathrm{SM}=(5.24\pm0.08)\%$. The SM prediction is calculated with $f_{D_s^+}=249.9\pm0.5$~MeV taken from the $N_f=2+1+1$ lattice result given by the  FLAG~\cite{FlavourLatticeAveragingGroupFLAG:2021npn}, and $|V_{cs}|=0.975\pm0.006$ quoted from the PDG~\cite{ParticleDataGroup:2022pth}. This branching ratio therefore provide us a stringent bound on the the Wilson coefficient $[C_{\ell q}^{(3)}]_{3323}$.

As for semileptonic charmed meson decays, we are interested in the dineutrino modes, i.e., $D\to P\nu\bar{\nu}$ decays. Since the relevant SM Wilson coefficient $C_{L,\mathrm{SM}}^{U\nu_\tau}$ is small, these processes are highly suppressed in the SM which turn out to be very sensitive to the NP. The differential branching ratio of $D\to P\nu\bar{\nu}$ decays can be written as
\begin{align}
\frac{d\mathcal{B}(D\to P\nu\bar{\nu})}{dq^2}=\tau_{D}\frac{G_F^2\alpha^2}{3072\pi^5m_D^3}f_{D\to P,+}^2(q^2)\lambda^{\frac{3}{2}}(q^2,m_D^2,m_P^2) \left|\Delta C_L^{U\nu_\tau}\right|^2\,,
\end{align}
which can be used to describe the $D^0\to\pi^0\nu\bar{\nu}$, $D^+\to\pi^+\nu\bar{\nu}$, and $D^+_s\to K^+\nu\bar{\nu}$ decays. After integrating over $q^2$ in the kinematic ranges (with $[0,(m_{D^0}-m_{\pi^0})^2]$ for $D^0\to\pi^0\nu\bar{\nu}$, $[0.34~\mathrm{GeV},(m_{D^+}-m_{\pi^+})^2]$ for $D^+\to\pi^+\nu\bar{\nu}$, and $[0.66~\mathrm{GeV},(m_{D_s^+}-m_{K^+})^2]$ for $D^+_s\to K^+\nu\bar{\nu}$~\cite{Bause:2020xzj,Bause:2020auq}, respectively), one obtains the following numerical expressions for the branching ratios:
\begin{align}
\mathcal{B}(D\to P\nu\bar{\nu})=A_+^{DP}\left|\Delta C_L^{U\nu_\tau}\right|^2\,,
\end{align}
where $A_+^{D^0\pi^0}=0.9\times10^{-8}$, $A_+^{D^+\pi^+}=3.6\times10^{-8}$, and  $A_+^{D_s^+K^+}=0.7\times10^{-8}$, respectively~\cite{Bause:2020xzj,Bause:2020auq}, with the form factors $f_{D\to P,+}(q^2)$ adopted from the lattice calculations of Ref.~\cite{Lubicz:2017syv}. Up to date, only a rough upper limit is reported for $D^0\to\pi^0\nu\bar{\nu}$ decay by the BESIII collaboration, with $\mathcal{B}(D^0\to\pi^0\nu\bar{\nu})_\mathrm{BESIII}<2.1\times10^{-4}$ at $90\%$ CL~\cite{BESIII:2021slf}, which is far higher than its SM prediction $\sim10^{-15}$~\cite{Burdman:2001tf}.
We will make predictions for these branching ratios with Wilson coefficient from the previous constraints.

\section{Numerical analysis}
\label{sec:num}

In this section we will use the experimental bounds of observables discussed in last section to constrain the relevant SMEFT Wilson coefficients listed in Eq.~\eqref{eq:SMEFT}, basing on the least squares method. Our strategy is to separate the total observables into two parts, one of which enters the $\chi^2$ function which is defined to constrain the parameters, while the remaining one will be used to test if the previous constraints are in agreements with the current experimental bounds. For convenience, the parameters inputs used in the numerical analysis throughout this work are collected in Table~\ref{tab:input}. One point we need to clarify is that the CKM matrix element $|V_{cb}|=(41.83_{-0.69}^{+0.79})\times10^{-3}$ is taken from the PDG~\cite{ParticleDataGroup:2022pth}, which is determined by using a global fit to all available measurements and imposing the unitarity constraints.

\begin{table}[h!]
\centering
\tabcolsep 0.20in
\begin{tabular}{|l|l|}
\hline
  $G_F = 1.1663788 \times 10^{-5}~\mathrm{GeV}^{-2}$  \hfill\cite{ParticleDataGroup:2022pth}  &
  $f_{B_s} = 227.7 \pm 4.5~\mathrm{MeV}$                  \hfill\cite{FlavourLatticeAveragingGroupFLAG:2021npn} \\
   $m_W = 80.377\pm0.012~\mathrm{GeV}$  
\hfill  \cite{ParticleDataGroup:2022pth} &
  $f_{B_c^+} = 427 \pm 6~\mathrm{MeV}$                  
\hfill\cite{Colquhoun:2015oha} \\
  $m_{\tau} = 1776.86\pm0.12~\mathrm{MeV}$            
\hfill \cite{ParticleDataGroup:2022pth} &
 $f_{B^+} = 190.0 \pm 1.3~\mathrm{MeV}$
\hfill \cite{FlavourLatticeAveragingGroupFLAG:2021npn}\\
  $m_{K^0} = 497.611\pm 0.013~\mathrm{MeV}$                
\hfill\cite{ParticleDataGroup:2022pth} &
$f_{D_s^+} = 249.9\pm 0.5~\mathrm{MeV}$
\hfill\cite{FlavourLatticeAveragingGroupFLAG:2021npn}\\
$m_{K^+} = 493.677 \pm 0.016~\mathrm{MeV}$        
 \hfill\cite{ParticleDataGroup:2022pth} &
  $\tau_{B_s} = 1.527 \pm 0.011~\mathrm{ps}$           \hfill\cite{ParticleDataGroup:2022pth}    \\
  $m_{K^{\ast 0}} = 895.55\pm 0.20~\mathrm{MeV}$
  \hfill\cite{ParticleDataGroup:2022pth} &
  $\tau_{B^0} = 1.519 \pm 0.004~\mathrm{ps}$           \hfill\cite{ParticleDataGroup:2022pth}    \\
  $m_{B_s} = 5366.77 \pm 0.24~\mathrm{MeV}$           
   \hfill\cite{ParticleDataGroup:2022pth}   &
 $\tau_{B^+} = 1.638 \pm 0.004~\mathrm{ps}$         \hfill\cite{ParticleDataGroup:2022pth}    \\
 $m_{B^+} = 5279.34 \pm 0.12~\mathrm{MeV}$           
   \hfill\cite{ParticleDataGroup:2022pth} &             
 $\tau_{B_c^+} = 0.510 \pm 0.009~\mathrm{ps}$         \hfill\cite{ParticleDataGroup:2022pth}    \\
 $m_{B_c^+} = 6274.47 \pm 0.32~\mathrm{MeV}$
 \hfill\cite{ParticleDataGroup:2022pth}   &
 $\tau_{D_s^+} = 0.504 \pm 0.004~\mathrm{ps}$        
 \hfill\cite{ParticleDataGroup:2022pth}\\
 $m_{B^0} = 5279.58 \pm 0.17~\mathrm{MeV}$
  \hfill\cite{ParticleDataGroup:2022pth} &
  $\tau_{D^+} = 1.033 \pm 0.005~\mathrm{ps}$
    \hfill\cite{ParticleDataGroup:2022pth}    \\
  $m_{D^0} = 1864.84 \pm 0.05~\mathrm{MeV}$
   \hfill\cite{ParticleDataGroup:2022pth} &
  $\tau_{D^0} = 0.4103 \pm 0.0010~\mathrm{ps}$            
  \hfill\cite{ParticleDataGroup:2022pth}    \\
    $m_{D^+} = 1869.66 \pm 0.05~\mathrm{MeV}$         
  \hfill\cite{ParticleDataGroup:2022pth} &  
   $|V_{ub}|=(3.70\pm0.11)\times10^{-3}$
 \hfill\cite{UTfit:2022hsi} \\
       $m_{D_s^+} = 1968.35 \pm 0.07~\mathrm{MeV}$       
  \hfill\cite{ParticleDataGroup:2022pth}   &
  $|V_{cb}|=(41.83_{-0.69}^{+0.79})\times10^{-3}$
  \hfill\cite{ParticleDataGroup:2022pth} \\
  $m_{\pi^0}=134.9768\pm0.0005~\mathrm{MeV}$
  \hfill\cite{ParticleDataGroup:2022pth}   &
  $|V_{tb}|=1.014\pm0.029$
  \hfill\cite{ParticleDataGroup:2022pth}\\
  $m_{\pi^+}=139.57039\pm0.00018~\mathrm{MeV}$
  \hfill\cite{ParticleDataGroup:2022pth}   &
  $|V_{ts}|=(41.5\pm0.9)\times10^{-3}$
  \hfill\cite{ParticleDataGroup:2022pth}\\
    $m_b(m_b) = 4.18 \pm 0.03~\mathrm{GeV}$                      
  \hfill\cite{ParticleDataGroup:2022pth} &  
  $|V_{cs}|=0.975\pm0.006$
  \hfill\cite{ParticleDataGroup:2022pth}\\
 $m_s(2~\mathrm{GeV}) = 93.4^{+8.6}_{-3.4}~\mathrm{MeV}$
 \hfill\cite{ParticleDataGroup:2022pth} & $A_+^{D^0\pi^0}=0.9\times10^{-8}$
 \hfill\cite{Bause:2020auq}\\
 $\alpha(4.18~\mathrm{GeV}) = 1/132.1 $ & $A_+^{D^+\pi^+}=3.6\times10^{-8}$\hfill\cite{Bause:2020auq}\\
$\alpha(2~\mathrm{GeV}) = 1/133.2$ & $A_+^{D_s^+K^+}=0.7\times10^{-8}$\hfill\cite{Bause:2020auq}\\
[0.15cm]
\hline
\end{tabular}
\caption{\small Relevant input parameters used in our numerical analysis.}
\label{tab:input}
\end{table}

\subsection{SMEFT Wilson coefficients}

Collecting the scattered informations discussed in last section, we are left with total four SMEFT Wilson coefficients that are needed to be constrained:
\begin{align}\label{eq:WCs}
\vec{\theta}=\left([C_{\ell q}^{(1)}]_{2223}\,, [C_{\ell q}^{(1)}]_{3323}\,, [C_{\ell q}^{(3)}]_{3323}\,,[C_{\ell d}]_{3323}\right)\,,
\end{align}
where the universal assumption $[C_{\ell q}^{(1)}]_{2223}=[C_{\ell q}^{(1)}]_{1123}$ and the deduction $[C_{\ell d}]_{1123}=[C_{\ell d}]_{2223}=0$ from the global fits of $b\to s\ell^+\ell^-$ in Ref.~\cite{Wen:2023pfq} are understood. These parameters enter the observables via matching onto the low-energy Wilson coefficients, as showed in Eqs.~\eqref{eq:match}. Assuming that these coefficients obey the normal distribution and following the general procedure of the least squares method, we can obtain the best-fit values by minimizing the following $\chi^2$ function:\begin{align}\label{eq:chisq}
\chi^2(\vec{\theta})=(\mathcal{O}_\mathrm{the.}(\vec{\theta})-\mathcal{O}_\mathrm{exp.})^\top V^{-1}(\mathcal{O}_\mathrm{the.}(\vec{\theta})-\mathcal{O}_\mathrm{exp.})\,,
\end{align}
where $\mathcal{O}_\mathrm{the.}$ and $\mathcal{O}_\mathrm{exp.}$ stand respectively for the theoretical expressions and measurements of the observables, and $V$ the corresponding covariance matrix encoding the total uncertainties, which are obtained by adding the experimental and theoretical ones in quadrature. The observables entering the $\chi^2$ function are as follows:
\begin{align}
\mathcal{O}=\left(\mathcal{B}(B^+\to K^+\nu\bar{\nu}),~\mathcal{B}(B^0\to K^{\ast0}\nu\bar{\nu}),~R_D,~R_{D^\ast},~\Delta C_9^{\mu}\right)\,.
\end{align}
Here, we assume that all of the experimental inputs of the observables are independent to each other except that a $R_D-R_{D^\ast}$ correlation of $-0.39$. Other observables that do not enter the $\chi^2$ function will be employed to cross check the consistency between the measurements and predictions, in which the latter are obtained with NP inputs from the results of fit. 

With all the numerical inputs and the theoretical expressions at hand, we can now obtain the final resulting constraints on the Wilson coefficients listed in Eq.~\eqref{eq:WCs}, which are given,
respectively, by
\begin{align}\label{eq:fit}
[C_{\ell q}^{(1)}]_{2223}=&(6.50^{+1.76}_{-1.73})\times10^{-4}\,, & [C_{\ell q}^{(1)}]_{3323}=&(5.57^{+1.34}_{-1.37})\times10^{-2}\,,\notag\\ 
[C_{\ell q}^{(3)}]_{3323}=&(4.75^{+1.15}_{-1.16})\times10^{-2}\,,& [C_{\ell d}]_{3323}=&(1.87^{+0.63}_{-0.74})\times10^{-2}\,,
\end{align}
with the best-fit values corresponding to a minimum $\chi^2_\mathrm{min}/\mathrm{d.o.f.}=1.22$, and the associated correlation matrix for these parameters is 
\begin{align}\label{eq:correlation}
\rho=\begin{pmatrix}
1 & -0.007 & 0 & -0.013\\
-0.007 & 1 & 0.846 & -0.285\\
0 & 0.846 & 1 & 0\\
-0.013 & -0.285 & 0 & 1
\end{pmatrix}\,.
\end{align}
To compare our constraints with that of other literatures, we collect the numerical results for the same group of Wilson coefficients obtained in different methods in Table.~\ref{tab:literatures}. Here all of the values correspond to the SMEFT Wilson coefficients at 1~TeV.

\begin{table}[h!]
\begin{center}
	\renewcommand\arraystretch{1}
		\resizebox{\textwidth}{!} 
{
			\begin{tabular}{lcccc}
			\hline\hline\noalign{\smallskip}
  Wilson coefficient & $[C_{\ell q}^{(1)}]_{2223}$ & $[C_{\ell q}^{(1)}]_{3323}$ & $[C_{\ell q}^{(3)}]_{3323}$ & $[C_{\ell d}]_{3323}$ \\  
 \hline
 This work & $(6.50^{+1.76}_{-1.73})\times10^{-4}$ & $(5.57^{+1.34}_{-1.37})\times10^{-2}$ & $(4.75^{+1.15}_{-1.16})\times10^{-2}$ & $(1.87^{+0.63}_{-0.74})\times10^{-2}$\\
HLY\cite{Hu:2018veh} & --- & --- & $-0.0679\pm0.0150$&---\\ 
 CFFPSV\cite{Ciuchini:2022wbq} & $[4.78, 9.33]\times10^{-4}$ & --- & ---&---\\
 Drell-Yan tails\cite{Greljo:2022jac} & $[-0.066, 0.071]$ & --- & ---&---\\
$b\to q\ell\ell$\cite{Greljo:2022jac} & $[7.71, 51.86]\times10^{-5}$ &--- &--- & ---\\
$b\to q\nu\nu$\cite{Greljo:2022jac} & $[-0.038, 0.017]$ & ---& ---& ---\\
$95\%$ CL LHC\cite{Allwicher:2022gkm} & $[-0.14,0.12]$ & $[-0.30,0.27]$ & $[-0.13,0.15]$ & $[-0.29,0.29]$\\
ABPRS muon-specific\cite{Allwicher:2023xba} & $[0.0129, 0.0134]$ &---&---&---\\
ABPRS flavor universal\cite{Allwicher:2023xba} & $[0.012, 0.013]$ & $[0.012, 0.013]$&---&---\\
ABPRS tau-specific\cite{Allwicher:2023xba} &---&---&$[-0.015, -0.013] \cup[0.033, 0.036]$&---\\
\hline
	\end{tabular}
		}
	\end{center}
	\caption{\small Comparisons of our constraints on the SMEFT Wilson coefficients with those of other references. The numerical values for all of these parameters are obtained at the scale 1~TeV.}
\label{tab:literatures}
\end{table}

\begin{figure}[h]
  \centering
  \includegraphics[width=\textwidth]{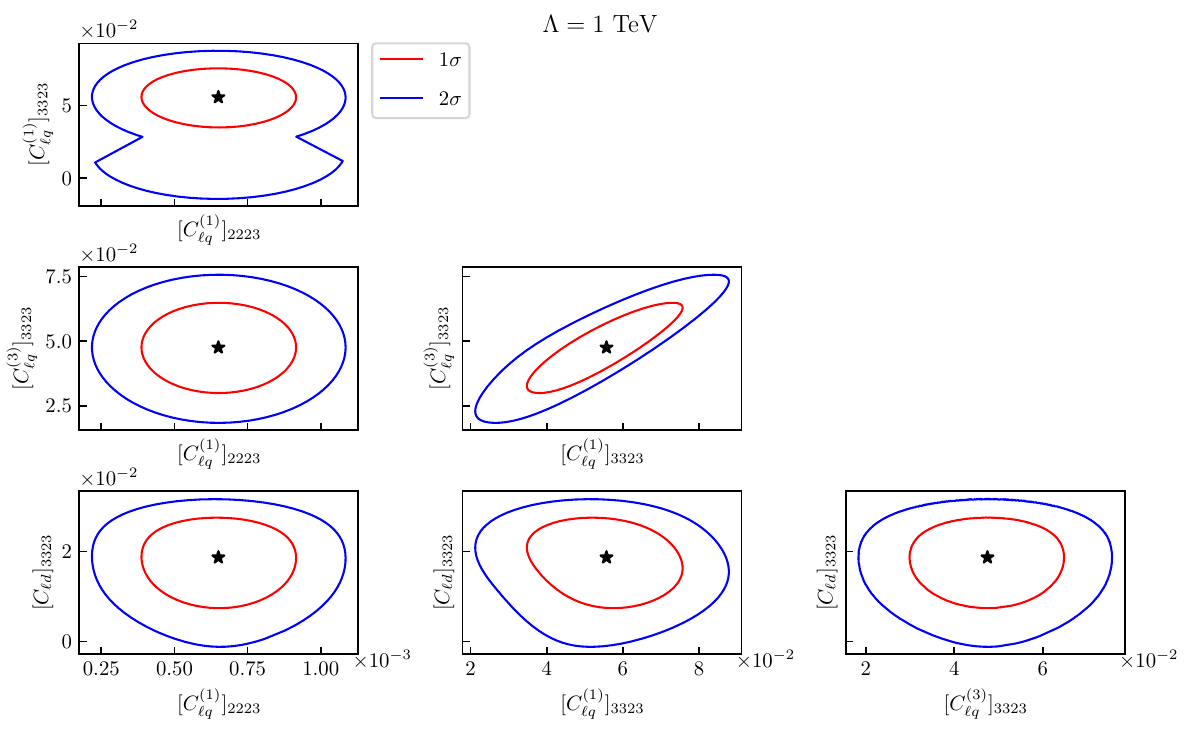}
    \vspace{-1.0cm}
  \caption{The the $1\sigma$ (red) and $2\sigma$ (blue) allowed regions for the fit results of the SMEFT Wilson coefficients $[C_{\ell q}^{(1)}]_{2223}$, $ [C_{\ell q}^{(1)}]_{3323}$, $ [C_{\ell q}^{(3)}]_{3323}$, and $[C_{\ell d}]_{3323}$ in Eq.~\eqref{eq:fit}, the asterisks stand for the center values of the best fit results. Here we have fixed the NP scale $\Lambda$ to be 1~TeV.}\label{fig:contour}
\end{figure}

It is clear from Eq.~\eqref{eq:correlation} that there exists a relative strong positive correlation ($0.846$) between $[C_{\ell q}^{(1)}]_{3323}$ and $[C_{\ell q}^{(3)}]_{3323}$, while the rest of correlations are rather weak. This can also be observed visually in the contour plots of parameters in Figure~\ref{fig:contour}, in which the $1\sigma$ and $2\sigma$ allowed regions of any two SMEFT Wilson coefficients are performed. As discussed in Sec.~\ref{sec:eft}, the NP scale $\Lambda$ in Eq.~\eqref{eq:fit} has been fixed to be 1~TeV. One can observe that the Wilson coefficient $[C_{\ell q}^{(1)}]_{2223}$ (so as $[C_{\ell q}^{(1)}]_{1123})$ is two orders of magnitude smaller than all of the couplings to the third generation of leptons, i.e., $ [C_{\ell q}^{(1)}]_{3323}$, $ [C_{\ell q}^{(3)}]_{3323}$, and $[C_{\ell d}]_{3323}$, which reflects a significant violation of the LFU. This implies that the latter can play a more important role in simultaneously explaining the discrepancies in the aforementioned rare $B$ decays. These observations also provide us the useful informations to build a feasible ultraviolet complete models.

\begin{table}[h!]
\centering
\tabcolsep 0.20in
\begin{tabular}{lll}
\hline\hline\noalign{\smallskip}
  Observable & prediction with NP & Experiment \\
 \hline
 $\mathcal{B}(B^+\to K^+\nu\bar{\nu})$ & $(2.06\pm0.69)\times 10^{-5}$ & $(2.3\pm0.7)\times 10^{-5}$ \hfill\cite{Belle-II:2023esi}\\
 $\mathcal{B}(B^0\to K^{\ast0}\nu\bar{\nu})$ & $(1.42\pm0.74)\times 10^{-5}$ & $<1.8\times10^{-5}$ \hfill\cite{Belle:2017oht}\\ 
 $R_D$ & $0.339\pm0.010$ & $0.344\pm0.026$ \hfill\cite{HFLAV:2022esi}\\
 $R_{D^\ast}$ & $0.289\pm0.009$ & $0.285\pm0.012$ \hfill\cite{HFLAV:2022esi}\\
 $\Delta C_9^{\mu}$ & $-0.782\pm0.212$ & $-0.789_{-0.210}^{+0.198}$\hfill\cite{Wen:2023pfq}\\
\hline
 $\mathcal{B}(B^+\to \tau^+\nu)$ & $(1.02\pm0.08)\times10^{-4}$ & $(1.09\pm0.24)\times10^{-4}$ \hfill\cite{ParticleDataGroup:2022pth}\\
 $\mathcal{B}(B_c^+\to \tau^+\nu)$ & $(2.64\pm0.23)\times10^{-2}$ &  \qquad---\\
 $\mathcal{B}(B_s\to \tau^+\tau^-)$ & $(4.26\pm2.85)\times10^{-4}$ &  $<6.8\times10^{-3}$\hfill\cite{LHCb:2017myy}\\
 $\mathcal{B}(D_s^+\to \tau^{+}\nu_\tau)$ & $(5.39\pm0.18)\times 10^{-2}$ & $(5.37\pm0.23)\times 10^{-2}$\hfill\cite{BESIII:2023ukh}\\
 $\mathcal{B}(D^0\to\pi^0\nu\bar{\nu})$ & $(2.21\pm1.05)\times10^{-11}$ & $<2.1\times10^{-4}$\hfill\cite{BESIII:2021slf}\\
 $\mathcal{B}(D^+\to\pi^+\nu\bar{\nu})$ & $(8.83\pm4.19)\times10^{-11}$&  \qquad---\\
 $\mathcal{B}(D_s^+\to K^+\nu\bar{\nu})$ &$(1.72\pm0.82)\times10^{-11}$ & \qquad--- \\
 $\langle F_L(B^0\to K^{\ast0}\nu\bar\nu)\rangle$ & $0.26\pm0.12$      & \qquad--- \\
\hline
\end{tabular}
\caption{\small Comparisons of the predictions with NP contributions and the experimental results. Observables above and below the horizontal line correspond to those ones enter and do not enter the $\chi^2$ function, respectively.}
\label{tab:comparison}
\end{table}

\begin{figure}[h!]
  \centering
  \includegraphics[width=\textwidth]{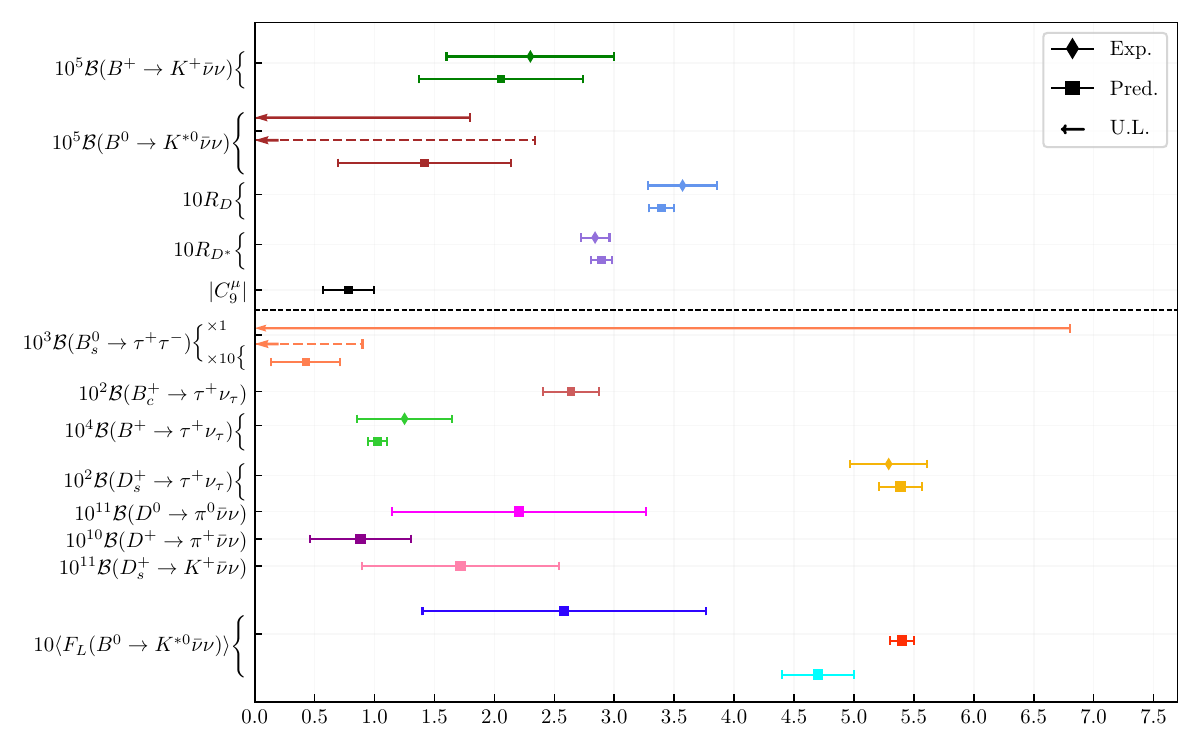}
  \vspace{-1.0cm}
  \caption{A visual comparisons of predictions with NP contributions and the experimental results. Observables above and below the dashed line correspond separately to those ones enter and do not enter the $\chi^2$ function, i.e., Eq.~\eqref{eq:chisq}. The solid and dashed arrows denote respectively the experimental and theoretical upper limits of the observables. The latter are translated from the predictions assuming that they obey the half-normal distribution. The red and cyan intervals denote the values of $\langle F_L(B^0\to K^{\ast0}\nu\bar\nu)\rangle$ quoted from Refs.~\citep{Altmannshofer:2009ma} and~\cite{Buras:2014fpa}, respectively.}\label{fig:reprediction}
\end{figure} 

\subsection{Predictions of Observables}

To see whether the observables of which either do or do not enter the $\chi^2$ function with NP contributions from our fit results obtained in Eq.~\eqref{eq:fit} are in agreements with the current experimental bounds, we shall substitute these parameters back into the theoretical expressions and remake predictions on them. For the convenience to have a clear comparison, we collect all the relevant observables in Table~\ref{tab:comparison}.\footnote{The observable $\Delta C_9^{\mu}$ is not a directly measurement but a global fit result taken from Ref.~\cite{Wen:2023pfq}.} As discussed previously, the observables in Table~\ref{tab:comparison} have been divided into two components with a horizontal line, in which the ones above and below the horizontal line correspond to those ones enter and do not enter the $\chi^2$ function, respectively. To visually show the comparisons of predictions and measurements for each of the observables, we also plot the corresponding informations of Table~\ref{tab:comparison} in Figure~\ref{fig:reprediction}, in which different observables have multiplied a different rescaling factors so that they can be depicted in a same axis. Several main features of Table~\ref{tab:comparison} are concluded as follows:
\begin{enumerate}
\item[(i)] For observables with nonzero measurements, our predictions with NP contributions are compatible with those of the experimental values within $1\sigma$. Specially, the prediction of $\Delta C_9^\mu$ is almost overlap with the one of the experiment. An important observation is that the deviations between the SM predictions and measurements on $\mathcal{B}(B^+\to K^+\nu\bar{\nu})$, $R_D$, and $R_{D^\ast}$, i.e., the so-called $B$ anomalies, can be simultaneously explained by the set of SMEFT Wilson coefficients listed in Eq.~\eqref{eq:WCs} with values in Eq.~\eqref{eq:fit}.
\item[(ii)] The prediction of $\mathcal{B}(B^0\to K^{\ast0}\nu\bar{\nu})$ is $(1.42\pm0.74)\times10^{-5}$, whose uncertainty is relative large, yet the allowed range is still compatible with the upper limit of experiment. Assuming that this prediction obeys the half-normal distribution, then one can translate its range into a formalism of upper limit (labelled as the brown dashed arrow in Figure~\ref{fig:reprediction}), which is slightly larger than the one of the current experiment (labelled as the brown solid arrow in the same figure). Therefore, observing such a decay mode in the future experiments is still promising. With the individual predictions on $\mathcal{B}(B^0\to K^{\ast0}\nu\bar{\nu})$ and $\mathcal{B}(B^+\to K^{+}\nu\bar{\nu})$ at hand, we obtain the following result polluted by NP for the ratio defined in Eq.~\eqref{eq:definition}: 
\begin{align}
R_\mathrm{NP}=0.69\pm0.43\,,
\end{align}
which is also compatible with the one of the current experiment bound $R_\mathrm{exp}=0.38\pm0.31$. As for the prediction of $\langle F_L(B^0\to K^{\ast0}\nu\bar\nu)\rangle$, the value in this work is smaller than the SM predictions given in Refs.~\citep{Altmannshofer:2009ma} and~\cite{Buras:2014fpa}, by a level of $2.3\sigma$ and $1.7\sigma$, respectively. The uncertainty of our prediction on $\langle F_L(B^0\to K^{\ast0}\nu\bar\nu)\rangle$ is still relatively large, the future developments of theoretical treatment as well as experimental measurement will help to improve our current prediction.

\item[(iii)] Similar observation also holds in $\mathcal{B}(B_s\to \tau^+\tau^-)$. Following the same steps as in (ii), one obtains a predicted upper limit for $\mathcal{B}(B_s\to \tau^+\tau^-)$ (labelled as the orange dashed arrow in Figure~\ref{fig:reprediction}), which is about one order of magnitude smaller than the current experiment bound that it may be tested by the future experiments like LHCb~\cite{Albrecht:2017odf} and CEPC~\cite{CEPCStudyGroup:2018ghi}.
\item[(iv)] The prediction of $\mathcal{B}(D_s^+\to \tau^{+}\nu_\tau)$ with NP contribution is in good agreement with the very recent measurement of BESIII~\cite{BESIII:2023ukh}.  
Similar observation also holds for $\mathcal{B}(B^+\to \tau^+\nu)$. The NP correction to $\mathcal{B}(B_c^+\to \tau^+\nu)$ is small, which turns out to that the prediction is consistent with the one of the SM at a level of $1.4\sigma$.
\item[(v)] The predictions of $\mathcal{B}(D\to P\nu\bar{\nu})$ is about four orders of magnitude larger than the ones of the SM, which, however, are still far small than the upper limit of the current experiment. Only improving the detection sensitivity by about seven orders with respect to the current one in the future can have the possibility to observe these decay modes.
\end{enumerate}

\section{Conclusion}
\label{sec:con}

In this study, we elucidate the observed $2.8\sigma$ deviation between the recent Belle II measurement and the SM prediction regarding $\mathcal{B}(B^+\to K^+\nu\bar{\nu})$, as well as the deviation in the ratio to $\mathcal{B}(B^0\to K^{\ast0}\nu\bar{\nu})$. This is achieved in a model-independent manner by leveraging effective field theories. Within the framework of LEFT, we characterize both $B^+\to K^+\nu\bar{\nu}$ and $B^0\to K^{\ast0}\nu\bar{\nu}$ at low energy through the same $b\to s \nu\bar{\nu}$ transition at the quark level.
Assuming that NP arises from a scale significantly above the EW scale and that the symmetry breaking in the SM is linearly realized, we derive pertinent LEFT operators from the $SU(3)_C\times SU(2)_L\times U(1)_Y$ invariant Lagrangian of the  SMEFT. Subsequently, we establish a connection between the LEFT and SMEFT, as the latter interfaces with other LEFT operators governing low-energy observables such as $R_{K^{(\ast)}}$, $R_{D^{(\ast)}}$, $\mathcal{B}(B_s\to\tau^+\tau^-)$, $\mathcal{B}(B_{(c)}^+\to\tau^+\nu_\tau)$, $\mathcal{B}(D_s^+\to\tau^+\nu\tau)$, $\mathcal{B}(D\to P\nu\bar{\nu})$, and more.
Conducting a comprehensive combined analysis, encompassing all relevant observables, we employ the least squares method to impose constraints. The numerical results for the SMEFT Wilson coefficients obtained from the $\chi^2$ fit are presented in Eq.~\eqref{eq:fit}. Utilizing these outcomes as inputs, we reevaluate predictions for the pertinent observables. The results, along with their comparisons to measurements, are numerically detailed in Table~\ref{tab:comparison} and visually depicted in Figure~\ref{fig:reprediction}. Notably, our predictions incorporating NP contributions from Eq.~\eqref{eq:fit} align well with the current experimental bounds within a $1\sigma$ range.

\section*{Acknowledgements}

We thank Xiao-Dong Ma for the useful discussion. This work is supported by NSFC under Grant Nos.~12475095 and U1932104, the Fundamental Research Funds for the Central Universities (11623330), and the 2024 Guangzhou Basic and Applied Basic Research Scheme Project for Maiden Voyage (2024A04J4190).

\noindent \textbf{Note Added.} All the authors contribute equally and they are co-first authors, while F. Xu is the corresponding author.


\bibliography{reference}

\end{document}